\documentclass[twocolumn,prb,superscriptaddress,showpacs,floatfix]{revtex4-1}
\usepackage[utf8]{inputenc}
\usepackage{mathtools}
\usepackage{amsmath}
\usepackage{amssymb}
\usepackage{graphicx}
\usepackage{esint}

\makeatletter

\providecommand{\tabularnewline}{\\}

\usepackage{float}
\usepackage[usenames]{color}
\usepackage{enumitem}
\usepackage{morefloats}
\usepackage{subfigure}
\usepackage{multirow}

\makeatother

\begin{document}

\title{Stochastic self-consistent Green's function second-order perturbation
theory (sGF2)}

\author{Daniel Neuhauser}

\email{dxn@ucla.edu}

\affiliation{Department of Chemistry and Biochemistry, University of California
at Los Angeles, California 90095 USA}

\author{Roi Baer}

\email{roi.baer@huji.ac.il}

\affiliation{Fritz Haber Center for Molecular Dynamics, Institute of Chemistry,
The Hebrew University of Jerusalem, Jerusalem 91904, Israel}

\altaffiliation{Roi Baer is on Sabbatical leave in the Department of Chemistry, University of California, Berkeley, California 94720, USA}

\author{Dominika Zgid}

\email{zgid@umich.edu}

\affiliation{Department of Chemistry, University of Michigan, Ann Arbor, Michigan
48109, USA}
\begin{abstract}
The second-order Green's function method (GF2) was shown recently
to be an accurate self-consistent approach for electronic structure
of correlated systems since the self-energy accounts for both the
weak and some of the strong correlation. The numerical scaling of
GF2 is quite steep however, $O(N^{5})$ (where the pre-factor is often
hundreds), effectively preventing its application to large systems.
Here, we develop a stochastic approach to GF2 (sGF2) where the self-energy
is evaluated by a random-vector decomposition of Green's functions
so that the dominant part of the calculation scales quasi linearly
with system size. A study of hydrogen chains shows that the resulting
approach is numerically efficient and accurate, as the stochastic
errors are very small, 0.05\% of the correlation energy for large
systems with only a moderate computational effort. The method also
yields automatically efficient MP2 energies and is automatically temperature
dependent. 
\end{abstract}
\maketitle
\begin{figure}
\centering \includegraphics[scale=0.25]{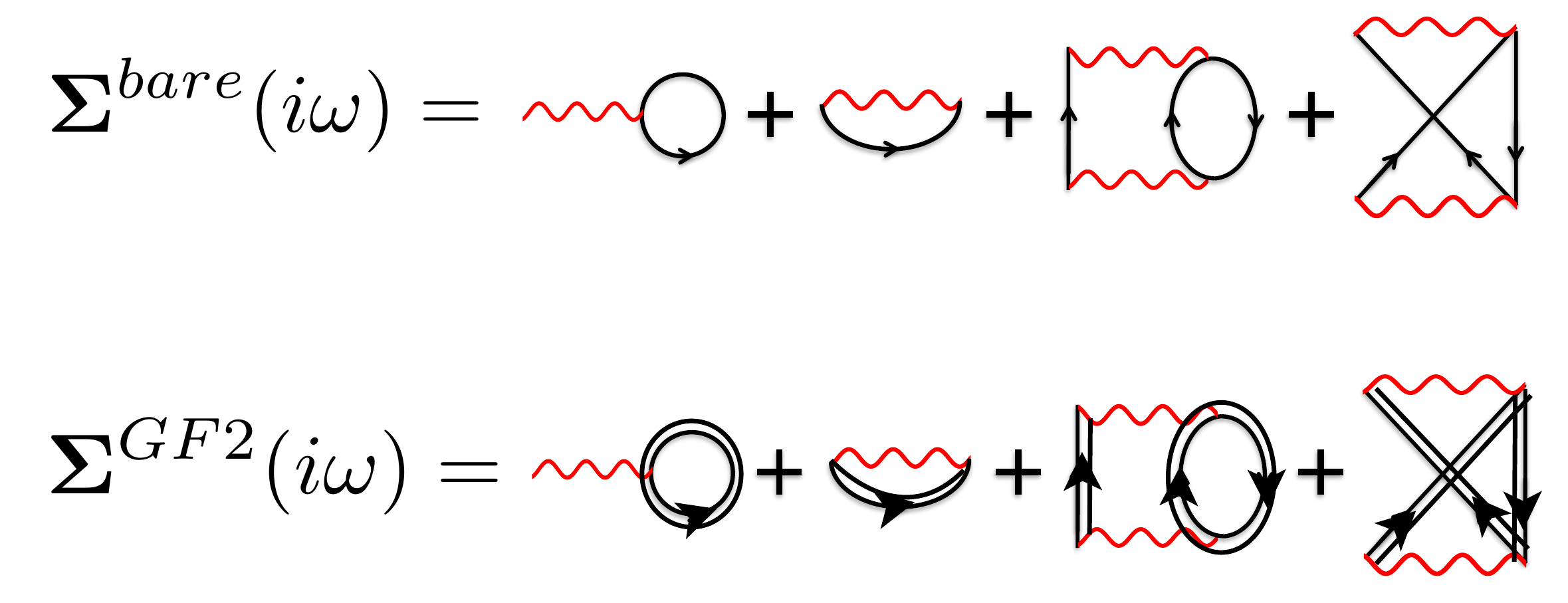} \caption{\label{SE_GF2}Bare and dressed second-order self-energy diagrams.}
\end{figure}

Second-order Green's function (GF2) is a temperature dependent self-consistent
perturbation approach where the Green's function is iteratively renormalized.
Thus, at self-consistency the self-energy which accounts for the many-body
correlation effects is a functional of the Green's function, $\Sigma(G)$.
Upon convergence the method includes all second order skeleton diagrams
dressed with the renormalized second order Green's function propagators,
as illustrated in Fig.~\ref{SE_GF2}. Specifically, as shown in Ref.~\onlinecite{Phillips:jcp/140/241101},
GF2, which at convergence is reference independent, preserves the
desirable features of Møller-Plesset perturbation theory (MP2) while
avoiding the divergences that appear when static correlation is important.
Additionally, GF2 possesses only a very small fractional charge and
spin error,\cite{Phillips:jcp/142/194108} less than either typical
hybrid density functionals or RPA with exchange, therefore having
a minimal self-interaction error. In solids~\cite{GF2_periodic_2016}
GF2 describes metallic, insulating, and Mott regimes and recovers
the internal energy for multiple phases present in a solid. Moreover,
GF2 is useful for efficient Green's function embedding techniques
such as in the self-energy embedding method (SEET).\cite{Kananenka:prb/91/121111,Lan_seet_2015}

While GF2 has these formal advantages, it is quite expensive and can
be easily applied only to small systems with up to a few hundred atomic
orbitals (AOs). This is because the calculation of the self-energy
matrix scales as $O(n_{\tau}N^{5})$, where $n_{\tau}$ is the size
of the imaginary time grid and $N$ the number of AOs. Here, we therefore
develop a statistical formulation which converts the many-indices
summation inherent in the self-energy matrix to a quasi linearly-scaling
approach.

The present development is in the same spirit as our previous work
on several electronic structure methods, including MP2,\cite{doi:10.1021/ct300946j,doi:10.1021/jz402206m}
RPA,\cite{doi:10.1021/jz3021606} DFT,\cite{doi:10.1021/ct500450w,:/content/aip/journal/jcp/141/4/10.1063/1.4890651}
DFT with long-range exact exchange,\cite{:/content/aip/journal/jcp/137/5/10.1063/1.4743959}
TDDFT,\cite{:/content/aip/journal/jcp/142/3/10.1063/1.4905568} and
GW.\cite{PhysRevLett.113.076402} Among these, the closest to this
work are the stochastic version of MP2 in real-time plane-waves,\cite{doi:10.1021/jz402206m}
and MO-based MP2 with Gaussian basis sets.\cite{doi:10.1021/ct300946j}
The present stochastic temperature dependent MP2 and GF2 are based
on a primary ingredient, the imaginary-time Green's function in an
AO basis, which is attractive for stochastic simulations since it
is smooth and varies continuously with parameter changes. We also
note other works which use stochastic sampling of perturbative quantum
chemistry expressions, including for example perturbative stochastic
approaches\cite{:/content/aip/journal/jcp/140/3/10.1063/1.4862255}
which employs different sampling schemes than ours, and a stochastic
coupled clusters algorithm (CC).\cite{PhysRevLett.105.263004}

The correlated Green's function is 
\begin{equation}
G(i\omega_{n})=\big[[G_{0}(i\omega_{n})]^{-1}-\Sigma(i\omega_{n})\big]^{-1},\label{eq:G}
\end{equation}
with $G_{0}(i\omega_{n})=\big[(\mu+i\omega_{n})S-F\big]^{-1},$ where
$S$ and $F$ are overlap and Fock matrices in the non-orthogonal
AO basis and $\mu$ is the chemical potential. The Matsubara frequencies
$\omega_{n}=(2n+1)\pi/\beta$ form a numerical grid where $\beta$
is the inverse temperature and $n$ is a non-negative integer. Recall
that the relation between quantities in Matsubara frequencies and
imaginary time is analogous to the relation between real-frequencies
and time, i.e., $G(i\omega_{n})=\int_{0}^{\beta}e^{i\omega_{n}\tau}G(\tau)dt$,
$G(\tau)=\frac{2}{\beta}{\rm {Re}}\sum_{n}e^{-i\omega_{n}\tau}G(i\omega_{n})$.
Further, an important feature is that $G(\tau)$ is a smooth function
of $\tau$. For example, $G_{0}(\tau)$ has a form

\begin{equation}
G_{0}(\tau)=\frac{e^{-\tau(\tilde{F}-\mu1)}}{1+e^{-\beta(\tilde{F}-\mu1)}},\label{eq:G0tau}
\end{equation}
where $\tilde{F}$ is the Fock matrix in an orthogonal basis and the
resulting $G_{0}(\tau)$ can be transformed to a non-orthogonal AO
basis if desired. Thus, $G_{0}(\tau)$ and $G(\tau)$ do not have
the oscillations of real-time Green's functions, and this is especially
important for the stochastic numerics below.

The Green's function is iteratively improved by updating the Fock-matrix
$F$ and the second order self-energy $\Sigma(i\omega_{n})$ that
describes the correlation effects. In GF2, the self-energy is 
\begin{eqnarray}
\Sigma_{ij}(\tau)=\sum_{klmnpq}G_{kl}(\tau)G_{mn}(\tau)G_{pq}(\beta-\tau)v_{ikmq}\label{eq:sigmat}\\
\times(2v_{ljpn}-v_{pjln}).\nonumber 
\end{eqnarray}

The Fock matrix here is $F=h+\Sigma_{\infty}$ where $h$ includes
the kinetic and nuclear-electron parts and $\Sigma_{\infty}$ is the
static, frequency independent part of the self-energy $[\Sigma_{\infty}]_{ij}=\sum_{kl}P_{kl}(v_{ijkl}-0.5v_{ilkj})$,
evaluated from the density matrix $P$ and the two-body integrals
$v_{ijkl}=\iint d\mathbf{r}d\mathbf{r}'\phi_{i}(\mathbf{r})\phi_{j}(\mathbf{r})|\mathbf{r}-\mathbf{r}'|^{-1}\phi_{k}(\mathbf{r}')\phi_{l}(\mathbf{r}')$,
assuming a real AO basis.

Typically, in the first iteration we obtain the density matrix $P$
from a Hartree-Fock (HF) solution; however, at convergence due to
the iterations GF2 is formally reference independent, so the initial
Fock-matrix can be DFT-based. Beyond the first iteration, the density
matrix is evaluated using the correlated Green's function as $P=\frac{2}{\beta}\sum_{n}e^{i\omega_{n}0^{+}}G(i\omega_{n})$
from which one constructs the Fock matrix.

Given the initial density matrix and Fock matrix, one iteratively
constructs (until self-consistency is reached) self-energy matrices
$\Sigma(i\omega_{n})$ and $\Sigma_{\infty}$, and the correlated
Green's function, density and Fock matrices. The chemical potential
$\mu$ is adjusted at each stage so that ${\rm {Tr}}(PS)$ yields
the correct number of electrons, $N_{e}$.

In GF2, both the Green's function and self-energy are smooth functions
of the imaginary time and frequency. At each stage, $G(i\omega_{n})$
is calculated and transformed to yield $G(\tau),$ which is used to
make the time-dependent self-energy $\Sigma(\tau)$ which is then
transformed to give $\Sigma(i\omega_{n})$. For computational efficiency,
we use a non-equidistant spline grid\cite{Kananenka:spline} sampling
the Matsubara frequencies. In the time domain, $\Sigma(\tau)$ is
expressed in an orthogonal polynomial basis.\cite{Kananenka:jctc/Leg}

The total energy is evaluated as: 
\begin{equation}
E=\frac{1}{2}{\rm {Tr}}\left[\left(h+F\right)P\right]+\frac{2}{\beta}\sum_{n}{\rm {Re}\big({\rm Tr}}[G(i\omega_{n})\Sigma^{T}(i\omega_{n})]\big),\label{eq:Etot}
\end{equation}
where the latter term can be also be also evaluated as $-\int{\rm {Tr}}\big(G(\tau)\Sigma(\beta-\tau)\big)d\tau$.
The GF2 correlation energy is defined as the difference between the
total energy and the Hartree-Fock energy, $E_{{\rm corr}}=E-E_{{\rm HF}}.$
Note that in the first iteration GF2 yields automatically the temperature-dependent
MP2 energy: 
\begin{equation}
E_{{\rm MP2}}=\frac{1}{\beta}\sum_{n}{\rm {Re}}\big({\rm Tr}[G_{HF}(i\omega_{n})\Sigma^{T}(i\omega_{n})]\big).\label{eq:Emp2}
\end{equation}

The main numerical challenge in GF2 is the determination of the time-dependent
self-energy (Eq.~\ref{eq:sigmat}), where, heuristically, two 4-index
tensors are connected through three matrices. To reduce the scaling
we turn to the stochastic paradigm which replaces matrices by a random
average over stochastically chosen vectors 
\begin{equation}
G_{kl}(\tau)=\big\{\eta_{k}(\tau)\bar{\eta}_{l}(\tau)\big\},\label{eq:Gtaurand}
\end{equation}
where the curly brackets refer to a stochastic average constructed
using random vectors. Here, we construct these vectors in the most
symmetric way possible, i.e., based on a square-root-like decomposition
of the real-symmetric matrix $G(\tau)$: 
\begin{equation}
\eta(\tau)=A(\tau)\sqrt{|g(\tau)|}A^{T}(\tau)\eta^{0}(\tau)\label{eq:etatau}
\end{equation}
\begin{equation}
\bar{\eta}(\tau)=A(\tau)\frac{g(\tau)}{\sqrt{|g(\tau)|}}A^{T}(\tau)\eta^{0}(\tau),\label{eq:etabar}
\end{equation}
where we introduced the eigenvectors and eigenvalues of the Green's
function, $G(\tau)=A(\tau)g(\tau)A^{T}(\tau)$, while $\eta^{0}(\tau)$
is a completely random vector, i.e., $\eta_{j}^{0}(\tau)=\pm1$ for
$j=1,...,N$. It is straightforward to prove Eq. \ref{eq:Gtaurand}
based on the fact that $\big\{\eta_{k}^{0}(\tau)\eta_{l}^{0}(\tau)\big\}=\delta_{kl}$.

We similarly separate the other two $G(\tau)$ matrices from Eq.~\ref{eq:sigmat},
writing them as $G_{mn}(\tau)=\big\{\zeta_{m}(\tau)\bar{\zeta}_{n}(\tau)\big\}$
and $G_{pq}(\beta-\tau)=\big\{\xi_{p}(\beta-\tau)\bar{\xi}_{q}(\beta-\tau)\big\}$.
The self-energy matrix becomes then an average over a separable product:
\begin{equation}
\Sigma_{ij}(\tau)=\big\{\bar{u}_{i}(\tau)(2u_{j}(\tau)-w_{j}(\tau))\big\},\label{eq:SigStoch}
\end{equation}
where formally $u_{i}(\tau)=\sum_{kmq}\eta_{k}(\tau)\zeta_{m}(\tau)\xi_{q}(\beta-\tau)v_{ikmq}$
with analogous expressions for $\bar{u},w$ (see below). It is efficient
to evaluate these summations on a grid, i.e., we define a time- and
space-dependent random function on a grid 
\begin{equation}
\eta(\mathbf{r},\tau)=\sum_{l}\eta_{l}(\tau)\phi_{l}(\mathbf{r}),\label{eq:etaRandGrid}
\end{equation}
with similar expressions for $\zeta(\mathbf{r},\tau)$, $\xi(\mathbf{r},\tau)$,
etc. (The computation of these functions is linear in system size
since the AO functions are local.) We then write the vectors decomposing
$\Sigma(\tau)$ as convolution integrals:

\begin{eqnarray}
u_{j}(\tau)=(\phi_{j}\eta(\tau)|\zeta(\tau)\xi(\beta-\tau))\quad\quad\quad\quad\quad\quad\quad\quad\quad\label{eq:ugrid}\\
\equiv\iint\phi_{i}(\mathbf{r})\eta(\mathbf{r},\tau)||\mathbf{r}-\mathbf{r}'|^{-1}\zeta(\mathbf{r}'\mathbf{,}\tau)\xi(\mathbf{r}',\beta-\tau)d\mathbf{r}d\mathbf{r}',\nonumber 
\end{eqnarray}
and analogously $w_{j}(\tau)=(\phi_{j}\xi(\beta-\tau)|\zeta(\tau)\eta(\tau)),$
$\ \bar{u}_{j}(\tau)=(\phi_{j}\bar{\eta}(\tau)|\bar{\zeta}(\tau)\bar{\xi}(\beta-\tau))$.
Therefore, the expensive summation over many indices is replaced by
averaging over convolution integrals, each of which scales quasi-linearly
with system size.

In passing, we note that the formalism is completely robust to near-degeneracies
in the basis due to the transformations in Eqs. \ref{eq:etatau},\ref{eq:etabar}.
This is in contrast to other formally simpler stochastic evaluations
of the self-energy, e.g., choosing randomly all the non$-ij$ indices
in Eq.\ref{eq:sigmat}.

{\em Computational details.} To study the performance of stochastic
GF2 we used typical model systems, finite (non-periodic) chains of
hydrogen atoms spaced 1 ${\rm \AA}$ apart with $N=$30, 100, 300
and 1000. A minimal STO-3G basis set was used so $N=N_{e}$. A numerical
grid contained up to 10$\times$10$\times$4000 points with a 0.5
bohr spacing was sufficient for converging Eq.~\ref{eq:ugrid}. The
convolution integrals employed the usual numerical-analytical splitting\cite{:/content/aip/journal/jcp/110/6/10.1063/1.477923}
which uses here grid-doubled integration due to the lack of periodicity
in any dimension. The inverse temperature was set at $\beta=50$ 1/a.u.
and a Chebyshev-type imaginary-time grid~\cite{Kananenka:jctc/Leg}
with 128 time points was employed. As mentioned, a spline-fit method
was used\cite{Kananenka:spline} for the frequency-to-time conversions
of $G(i\omega_{n})$ and $\Sigma(i\omega_{n})$ and for the evaluation
of the two-body energy.

The completely-random vectors ($\eta^{0}$,$\zeta^{0},\xi^{0}$) were
different at each time point $\tau$ but did not change with the self-consistent
GF2 iterations. A key for the self-consistency convergence was DIIS
(direct inversion of the iterative subspace).\cite{Pulay1980393,JCC:JCC540030413}
After each iteration a combined vector made from the $\tau$-dependent
self-energy $\Sigma(\tau)$ and the Fock matrix (weighted by a factor,
typically 3.0) was fed into a general DIIS routine that held in memory
the 4 previous iterations and updated the combined vector. The simulations
started with Hartree-Fock densities, and about 12 iterations were
required for the DIIS convergence to an error which was much smaller
than the statistical (sampling) error. The number of stochastic samples,
$N_{MC},$ ranged between 50 and 800, and the results were averaged
over several simulations starting from different random numbers.

\begin{figure}
\centering \includegraphics[scale=0.35]{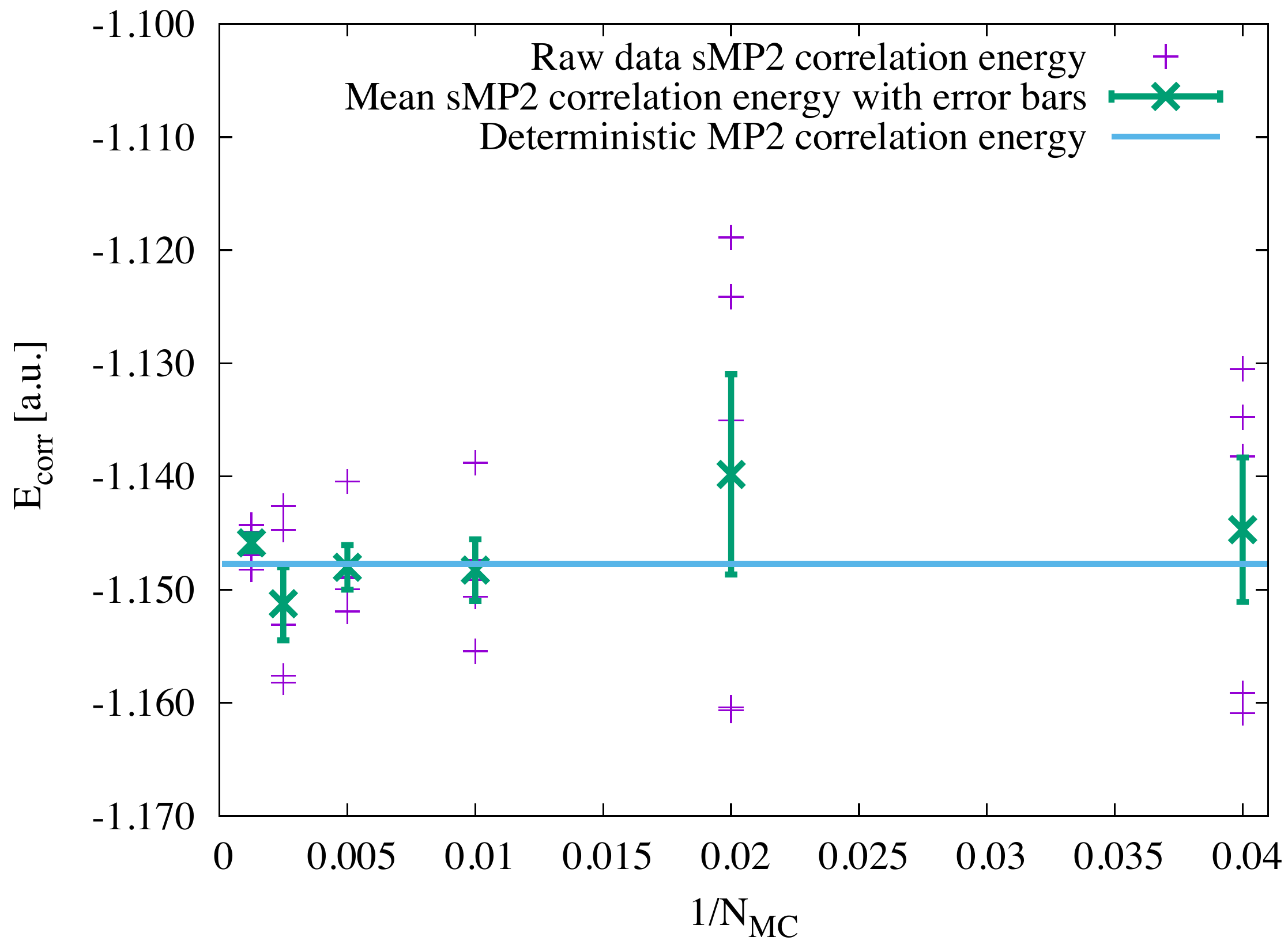} \caption{\label{E_MP2_100}The stochastic and deterministic thermal MP2 correlation
energies for different $N_{MC}$ for hydrogen chain with N=100.}
\end{figure}

{\em Temperature-dependent MP2.} The first SGF2 iteration yields
the MP2 correlation energy (Eq.~\ref{eq:Emp2}) which follows a Gaussian
distribution as illustrated in Fig.~\ref{E_MP2_100}. Table~\ref{tab:Convergence-with-}
shows the correlation energy $E_{{\rm corr}}$ and the (stochastic)
error $\delta E_{{\rm corr}}$ for chains with various number of orbitals.
Even for $N=1000$, the stochastic error is only 0.058\% of the total
correlation energy. For perspective, note that (deterministic) errors
of larger or similar magnitude are present in local or divide and
conquer MP2 methods with density fitting~\cite{:/content/aip/journal/jcp/118/18/10.1063/1.1564816,:/content/aip/journal/jcp/144/5/10.1063/1.4940732}.

\begin{figure}
\centering \includegraphics[scale=0.35]{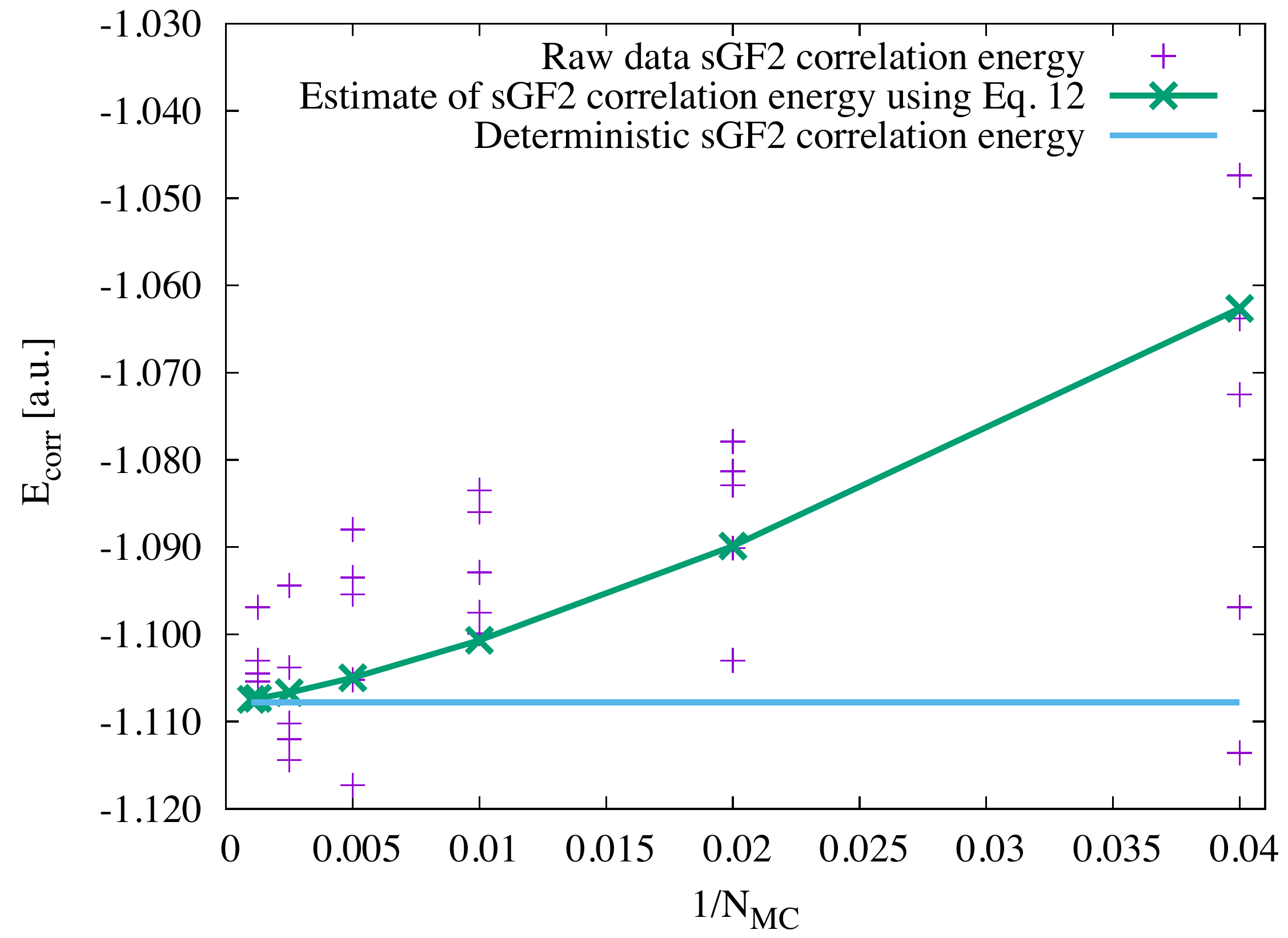} \caption{\label{E_GF2_100}The stochastic and deterministic GF2 correlation
energies for different $N_{MC}$ for hydrogen chain with N=100.}
\end{figure}

\begin{figure}
\centering \includegraphics[scale=0.35]{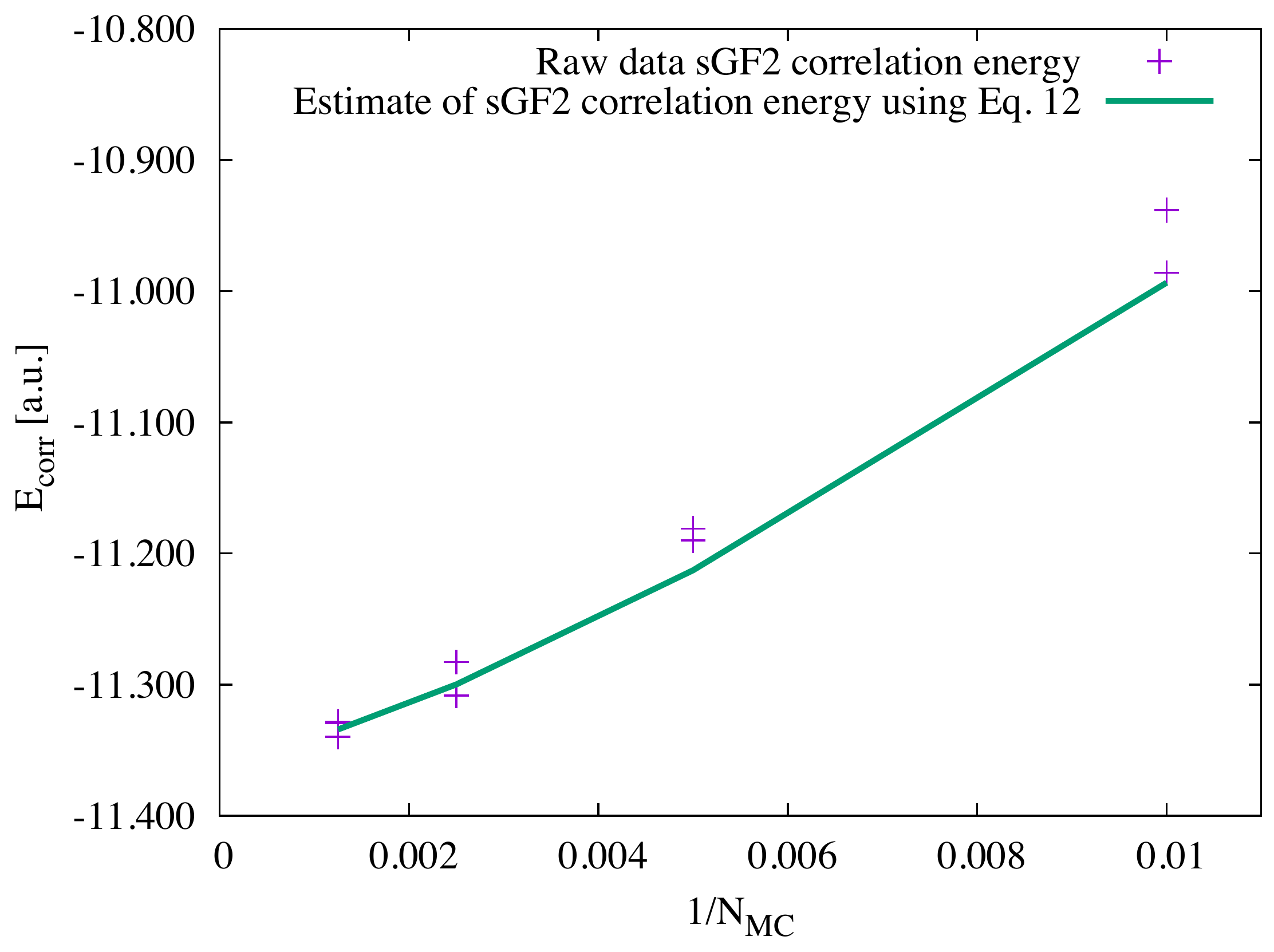}

\caption{\label{E_GF2_1000}The stochastic GF2 correlation energies for different
$N_{MC}$ for hydrogen chain with N=1000.}
\end{figure}

{\em Stochastic GF2.} In sGF2, due to the iterations, the dependence
of the energy on the Green function, which depends itself on the number
($N_{MC}$) of stochastic sampling of Eq.\ref{eq:SigStoch}, leads
to an additional non-stochastic (bias) dependence on $N_{MC}.$ (Simply
put, averaging a million sGF2 calculations each with 100 stochastic
samples will give different results than averaging 100 simulations
which each uses a million samples, unlike in sMP2.) Mathematically,
the energy in any specific simulation with $N_{MC}$ samplings is
then approximately $E=E(N_{MC})+r(N_{MC})^{-\frac{1}{2}}\delta$,
where $E(N_{MC})$ is a bias-type component, $r$ is a random scale-less
variable, and $\delta$ is approximately independent of $N_{MC}.$
Specifically, as seen in Figs.~\ref{E_GF2_100} and \ref{E_GF2_1000},
the energies for different $N_{MC}$ are stochastically distributed
around a line $E(N_{MC})$. We found excellent fit to an empirical
formula 
\begin{equation}
E(N_{MC})-E_{{\rm HF}}=E_{{\rm corr}}+b(N_{MC})^{-\frac{4}{3}}\label{eq:ENMC}
\end{equation}
i.e., the stochastic calculated energies cluster near the line defined
by this formula. We therefore fit, for any given system, the values
of $b$, and $E_{{\rm corr}}$ that yield the best fit to the actual
calculated data. The fitted value of $E_{{\rm corr}}$ is the GF2
prediction for the correlation energy. Note that an alternative is
to reduce the dependence of the curve (i.e., to flatten the $E(N_{MC})$
bias-dependence on $N_{MC}$) by employing in Eq. \ref{eq:Etot} Green's
functions and self-energies that come from different sets of stochastic
runs, in the spirit of ``jackknife'' analysis~\cite{Troyer}.

The sGF2 results, summarized in Table~\ref{tab:Convergence-with-},
show that the self-consistent iterations that are inherent in GF2
do not destroy the convergence of the method. Further, in large systems
the GF2 correlation reduces the spatial extent of $\Sigma_{\infty}$
so the statistical error even decreases relative to sMP2. The statistical
error in the total energy would have grown as $\sqrt{N}$ if the system
was made from completely isolated non-interacting fragments, and the
results here overall follow this trend.

\begin{table}
\begin{centering}
\begin{tabular}{|c|c|r|r|r|}
\hline 
Method  & $N$  & $E_{{\rm corr}}$  & $\delta E_{{\rm corr}}$  & $b$\tabularnewline
\hline 
\hline 
\multirow{4}{*}{sGF2}  & 30  & -0.307  & 0.0013  & -0.9\tabularnewline
\cline{2-5} 
 & 100  & -1.104  & 0.0021  & 3.3\tabularnewline
\cline{2-5} 
 & 300  & -3.388  & 0.0032  & 12.0\tabularnewline
\cline{2-5} 
 & 1000  & -11.357  & 0.0057  & 168.7\tabularnewline
\hline 
\multirow{4}{*}{sMP2}  & 30  & -0.337  & 0.0009  & \multicolumn{1}{r}{}\tabularnewline
\cline{2-4} 
 & 100  & -1.148  & 0.0016  & \multicolumn{1}{r}{}\tabularnewline
\cline{2-4} 
 & 300  & -3.472  & 0.0027  & \multicolumn{1}{r}{}\tabularnewline
\cline{2-4} 
 & 1000  & -11.610  & 0.0067  & \multicolumn{1}{r}{}\tabularnewline
\cline{1-4} 
\end{tabular}
\par\end{centering}

\raggedright{}\protect\caption{Stochastic GF2 and MP2 correlation energies in Hartree using 4000
total Monte Carlo runs (composed of several sets of runs with different
values of $N_{MC}$) for different number of orbitals in hydrogen
chains. For each system, an empirical formula $E-E_{HF}=E_{{\rm corr}}+b\left(N_{MC}\right)^{-\frac{4}{3}}$
was fitted to the calculated energies, so $E_{{\rm corr}}$ is the
predicted GF2 correlation energy.}
\label{tab:Convergence-with-} 
\end{table}

{\em Timings.} Overall the calculations took, for the largest system
size ($N_{e}=1000)$, 6500 core hours on standard 2Gflops CPUs, and
the method is fully parallelized. The temperature-dependent MP2 evaluations
only require the first (of twelve) SCF iterations so they took only
500 CPU core hours for the largest system. The calculation times are
dominated by the stochastic evaluation of the self-energy which rises
linearly with system size. For finite systems beyond $\approx5000$
orbitals the quadraticly and cubicly rising parts in the non-stochastic
GF2 parts (e.g., inversion and diagonalization of the Green's functions,
etc.) would start dominating the scaling.

In conclusion, we presented an sGF2 algorithm and evaluated electronic
correlation energies for prototypical hydrogen chain systems. We demonstrated
that the stochastic GF2 errors are well controlled and highly accurate
correlation energies with only a 0.05\% stochastic error are reached
with reasonable computational effort. These errors are comparable
with errors of local MP2 approaches used in quantum chemistry. The
present calculations are among the first applications of fully self-consistent
Green's function method with full imaginary frequency dependence to
large systems described by a full quantum chemistry Hamiltonian. Moreover,
we demonstrated that the splitting of matrices by a random average
over stochastically chosen vectors leads to a small variance and that
relatively few Monte Carlo samples already yield quite accurate correlation
energies. The reason for this is two-fold: the stochastic sampling
inherently acts only in the space of orbitals but the actual spatial
integrals (Eq. \ref{eq:ugrid}) are evaluated exactly; in addition,
since the Green's function matrices are smooth in imaginary time,
different random vectors can be used at each imaginary-time point
thereby enhancing the stochastic sampling efficiency.

There are many possible applications and extensions of the approach
studied here. sGF2 is potentially suitable for calculating energy
differences between neighboring configurations because the underlying
stochastic vectors are chosen automatically in the AO space so they
are continuous with a change of the system parameters; this is in
contrast with our previous stochastic MP2 basis in Gaussian basis
study\cite{doi:10.1021/ct300946j} where the stochastic vectors were
chosen in the MO space so they were not continuous with change of
parameters.

Further, the present stochastic GF2 and MP2 methods are automatically
suitable for periodic systems, as all the deterministic steps (i.e.,
Eq. \ref{eq:G}) and the time-frequency transforms are very efficient
when done in the reciprocal $(k)$ space. The only caveat is that
in periodic systems one needs to choose the random vectors to be in
$k$-space and then to convert them to real-space, as detailed in
an upcoming article.

Finally, we also note that, beyond the results presented here, it
should also be possible to achieve further reduction of the stochastic
error with an embedded fragment approach similar to self-energy embedding
approaches where a deterministic self-energy is embedded into the
stochastic estimate of self-energy for a larger fragment.

Discussions with Eran Rabani are gratefully acknowledged. D.N. and
D.Z. were supported by the NSF, grants CHE-1112500 and CHE-1453894,
respectively. D.N. and R.B. were also supported by the BSF grant 2012050.
R.B. is supported by the Israel Science Foundation--FIRST Program
(Grant No. 1700/14), and is supported for his sabbatical visit by
the Pitzer Center and the Kavli Institute of the University of California,
Berkeley. \bibliographystyle{jcp}
\bibliography{refs}

\end{document}